\documentstyle[preprint,prl,aps]{revtex}
\begin{document}
\draft

\title{Gauge drag between half-filled Landau levels}

\author{Yong Baek Kim$^a$ and A. J. Millis $^b$}
\address{$^a$Bell Laboratories, Murray Hill, NJ 07974\\
$^b$ Department of Physics and Astronomy, The Johns Hopkins University, 
Baltimore, MD 21218}

\date{November 14, 1996}

\maketitle

\begin{abstract}

The transresistance ({\it i.e.} the voltage induced in one
layer by a current in another) between composite fermions in 
double-layers of half-filled Landau levels is shown to be
dominated by scattering due to singular gauge field fluctuations 
arising from the antisymmetric combination of density 
fluctuations in two layers. 
The drag rate is found to be proportional to
$T^{4/3}$ for $T$ less than an energy scale
$\Omega_{\rm cr}$ which depends on the spacing, $d$, 
between the layers.
An observation of a $T^{4/3}$ temperature dependence
in the transresistance would be a strong evidence
for the existence of the gauge field fluctuations.

\end{abstract}
\pacs{PACS numbers:}

Recently there has been much interest in possible non-Fermi-liquid 
behavior of strongly correlated electron systems, stimulated in 
part by the unusual properties of high temperature 
superconductors\cite{Highsuper}. 
One of the characteristics which distinguishes these
systems from the usual Fermi-liquid is an anomalous temperature
dependence of the transport properties corresponding to
a scattering rate of $T^{\alpha}$ with $\alpha < 2$ instead 
of the usual $T^2$ dependence of a Fermi liquid.
One interesting class of proposed explanations involves 
scattering off a novel soft collective gauge field mode\cite{Nagaosa}.

A closely related system is the two dimensional electron gas in 
a high magnetic field at the filling fraction $\nu=1/2$, which has
been found experimentally to be a compressible liquid displaying
metallic properties\cite{Experiments}.
Using the Chern-Simons gauge transformation and the concept
of the composite fermions\cite{Jain,Lopez,Kalmeyer}, 
Halperin. Lee, and Read constructed
a theory for the $\nu =1/2$ state and other even denomenator 
filling fractions\cite{HLR}. 
In their theory, a Chern-Simons gauge transformation is introduced 
in which an even number of flux
quanta are attached to each electron making a composite fermion.
The composite fermion moves in a field given by the sum of 
the external magnetic field and the Chern-Simons statistical 
magnetic field\cite{Kalmeyer,HLR}.
In the mean field theory  at even-denomenator filling fractions,
the sum of these two magnetic fields vanishes so that
the composite fermions see zero magnetic field 
and form a Fermi sea which may explain the experimental 
observations. 

Corrections to the mean field theory involve fluctuations 
of the Chern-Simons gauge field; these turn out to be strong 
enough to generate a singular self-energy correction which invalidates 
the existence of the quasiparticle in the sense of the 
Landau-Fermi-liquid theory\cite{HLR,Kim1,Ioffe,Kim2,Kim3,Kim4,Stern}. 
Thus the system can be regarded as an example of 
non-Fermi-liquid\cite{Kim3,Kim4,Stern,Theory}. 
However, in the conventional experimental geometry for a single layer, 
the long-range Coulomb interaction suppresses the effects of the 
gauge field fluctuations. It turns out that the self-energy correction
is $\Sigma (i\omega) \propto i\omega {\rm ln} \omega$ so that the 
effective mass is only logarithmically singular. Thus, for a single
layer in the conventional geometry, the non-Fermi-liquid
behavior is marginal and not easily detected
experimentally.

A double-layer system of half-filled Landau levels is different;
it supports two distinct collective modes or
gauge fields corresponding to the symmetric and 
the antisymmetric combinations of density fluctuations in two layers\cite{Bonesteel}. 
Because the antisymmetric density fluctuations are screened at
sufficiently low energy, they lead to the stronger self-energy 
correction $\Sigma (i\omega) \propto \omega^{2/3}$ while the 
symmetric gauge field fluctuation gives $\Sigma (i\omega) 
\propto i\omega {\rm ln} \omega$\cite{Bonesteel}.
Thus one can expect that the double layer system shows 
stronger and more easily observable non-Fermi-liquid properties.

Double-layer systems also allow a novel measurement of 
the scattering mechanism\cite{Eisenstein}.
If (as is often the case) there is no tunneling between two layers, 
momentum can be transfered from one layer to another only via 
a scattering event between a carrier in one system
and a carrier in the other system. 
As a result, if a current is driven through one of the subsystems 
(active layer), then an induced current is dragged in the other 
system (passive layer) and the magnitude of the current is a measure
of the scattering rate.
In real experiments, no current is allowed to flow in
the passive layer so that a voltage is induced.
The ratio between the induced voltage in the passive layer
and the driven current in the active layer is the 
so-called transresistance which has been measured
for Coulomb-coupled double-quantum-well systems\cite{Eisenstein},
electron-hole systems\cite{Sivan}, and normal-metal-superconductor 
systems\cite{Giordano}. The transresistance for each system has 
also been calculated theoretically\cite{Jari,Oreg,MacDonald}.
In this paper, we calculate the temperature dependence
of the transresistance in double layers of 
half-filled Landau levels. 

Let us consider a double-layer electron system in a
high magnetic field.  We write the layer spacing as $d$,
the electron density in one layer as $n_e$, and the magnetic field as $B$.
We shall suppose that the filling fraction $\nu$ is an
even denominator fraction such as $1/2$ or $1/4$ for which a
compressible state would be formed in an isolated layer.
We define the
quantity ${\tilde \phi}$ from the equation
$B = {\tilde \phi} n_e \Phi_0$ with 
$\Phi_0 = hc/e$ being the flux quantum.  In this paper
we will use $\hbar = e = c =1$ so $\Phi_0 = 2 \pi$.
For the principal series $\nu = 1/2m$, ${\tilde \phi} = 2m$.  
It is also convenient to introduce the magnetic 
length $l_B = (B/(2\pi))^{1/2}$ and the Fermi wavevector $k_F$
of the compressible state, where $k_F = \sqrt{4 \pi n_e}
= 1/(\sqrt{m} l_B)$. In particular, at $\nu = 1/2$, 
${\tilde \phi} = 2$ and $k_F = 1 / l_B$.

We assume that all the spins are polarized and there is 
no tunneling between two layers.
When the layer spacing $d$ is comparable to or shorter than the 
interparticle spacing in one plane, the interlayer Coulomb
interaction becomes as strong as the intralayer one and 
the ground state of the system is expected to
be a quantum Hall liquid\cite{Halperin}. 
One the other hand, if $d$ is sufficiently larger than the 
interparticle spacing, 
each layer can be considered as a compressible
$\nu=1/2$ state at least in the physically relevant range of 
temperatures\cite{Bonesteel2}. 

To study the case of well separated layers, one can perform two
independent flux attachment transformations for electrons 
in different layers and introduce corresponding two 
Chern-Simons transverse gauge fields, 
${\bf a}^{\alpha}$ ($\alpha = 1,2$).
Here the bold face quantities are vectors in two spatial 
dimensions.
The low energy effective action for the system can be
written as
\begin{eqnarray}
S &=& \int d^2 r \ d\tau \ {\cal L} \ , \cr
{\cal L} &=& \psi^*_{\alpha} (\partial_{\tau} - 
i a^{\alpha}_0) \psi_{\alpha} + 
{1 \over 2 m^*} \psi^*_{\alpha} (-i\nabla + 
{\bf a}^{\alpha} - e {\bf A})^2 \psi_{\alpha} \ , \cr
&&- {i \over 2 \pi {\tilde \phi}} a^{\alpha}_0
\nabla \times {\bf a}^{\alpha} +
{1 \over 2 (2 \pi {\tilde \phi})} \int d^2 r'
\rho_{\alpha} ({\bf r},\tau)
v_{\alpha \alpha'} ({\bf r} - {\bf r'})
\rho_{\alpha'} ({\bf r'},\tau) \ ,
\label{lagrangian} 
\end{eqnarray}
where $\psi_{\alpha}$ and $\rho_{\alpha} = 
\psi^*_{\alpha} \psi_{\alpha}$ represent the composite 
fermion operator and the density operator in each layer 
$\alpha$ respectively. 
$B = \nabla \times {\bf A}$ is the
external magnetic field and the Coulomb gauge 
$\nabla \cdot {\bf a}^{\alpha} = 0$ is chosen for the
Chern-Simons gauge fields.
The effective mass, $m^*$, is given by the interaction energy 
scale in the lowest Landau level ($k^2_F / 2m^*$ is of the order 
of $e^2 / \varepsilon l_B$) and it is estimated as
$1/m^* = 0.3 e^2 l_B / \varepsilon$\cite{HLR}.
The constraints $\nabla \times {\bf a}^{\alpha} = 
2 \pi {\tilde \phi} \ \psi^*_{\alpha} \psi_{\alpha}$ are 
incorporated in the action through a Lagrange multiplier
which may be interpreted as the time component $a^{\alpha}_0$
of the gauge field.
$V_{\alpha \alpha'}$ represents the Coulomb interaction 
between the composite fermions in layer $\alpha$ and
the composite fermions in layer $\alpha'$; it is given by
\begin{equation}
V_{\alpha \alpha'} ({\bf r}) =
{2 \pi e^2 \over \varepsilon \sqrt{r^2 + 
d^2 (1 - \delta_{\alpha \alpha'})}} \ ,
\label{coulomb}
\end{equation}
where $\varepsilon$ is the dielectric constant.

The mean field approximation corresponds to taking the average of 
the Chern-Simons statistical magnetic field. In this case, the 
constraints are replaced by 
$\langle \nabla \times {\bf a}^{\alpha} \rangle = 
2 \pi {\tilde \phi} n_e$.
At the filling fraction $\nu=1/2$ per layer 
(${\tilde \phi} = 2$), there is a cancellation between the
external magnetic field and the averaged statistical 
magnetic field so that the composite fermions in each
layer form a Fermi sea with a Fermi wave vector
$k_F = \sqrt{4 \pi n_e}$. 
Corrections to the mean field theory involve fluctuations
of the gauge field $a_{\mu}$. It turns out that 
the longitudinal gauge fluctuations are not singular 
enough to generate an anomalous scattering rate. 
Thus we consider only the transverse gauge field 
fluctuations.    
There are two transverse gauge field fluctuations which we write as
${\bf a}^{\pm} = ({\bf a}^1 \pm {\bf a}^2) / \sqrt{2}$ 
corresponding to  symmetric 
and the antisymmetric combinations of the gauge field fluctuations.
The propagators $D^{\pm} ({\bf q},i\nu)$ 
in the RPA are given by
\begin{equation}
D^{\pm} ({\bf q},i\nu) \approx {1 \over \gamma |\nu|/q + 
\chi_0 q^2 + {q^2 \over (2 \pi {\tilde \phi})^2 m^*}
[2 \pi / m^* + V_{11} ({\bf q}) \pm V_{12} ({\bf q})]} \ ,
\end{equation}
where $\gamma = 2n_e / k_F$ and $\chi_0 = 1 / 24 \pi m^*$.  
If $q \ll 1/d$ and $q \ll k_F$, the propagators can be 
approximated as\cite{Bonesteel}
\begin{eqnarray}
D^+ ({\bf q},i\nu) &\approx& {1 \over 
\gamma |\nu|/q + \chi^+ q} \cr
D^- ({\bf q},i\nu) &\approx& {1 \over 
\gamma |\nu|/q + \chi^- q^2} \ ,
\label{propa}
\end{eqnarray}
where $\chi^+ = e^2 / \pi \varepsilon ({\tilde \phi})^2$ and 
$\chi^- = \chi_0 + 
(1 + e^2 m^* d / \varepsilon) / 2 \pi {\tilde \phi}^2 m^*$.
On the other hand, for $q \gg 1/d$, $D^+ \approx D^-$.
The ${\bf a}^-$ fluctuation is seen from Eq.~\ref{propa}
to be more infrared singular than ${\bf a}^+$ in the limit 
$qd \ll 1$.

The conductivity tensor $\sigma^{\alpha \beta}$
can be calculated in terms of
the retarded current-current correlation function 
$\Pi^{\alpha \beta, R}$.
We are interested in $\sigma^{12}$ which gives the current induced
in plane 1 by an electric field in plane 2.  This may be expressed as
\begin{equation}
{\rm Re} \ \sigma^{12} ({\bf q},\Omega) 
= {i \over \Omega} \Pi^{12, R}_{yy} 
({\bf q},\Omega) \ ,
\end{equation}
(We have assumed, without loss of generality, that ${\bf q}$ is parallel 
to ${\hat {\bf x}}$).
The retarded current-current correlation function
$\Pi^{\alpha \beta, R}_{yy} ({\bf q}, \Omega)$ is
the Fourier transform of  
\begin{equation}
\Pi^{\alpha \beta, R}_{yy} ({\bf r}-{\bf r'}, t-t')
= - i \Theta (t-t') 
\langle [ j^{\alpha}_y ({\bf r},t), j^{\beta}_y
({\bf r'},t') ] \rangle \ .
\end{equation}
$\Pi^{12, R}_{yy} ({\bf q},\Omega)$ will
be calculated from the analytic continuation
$i\nu \rightarrow \Omega + i\delta$ of the current-current 
correlation function $\Pi^{12}_{yy} ({\bf q},i\nu)$ 
in Euclidean space.

The lowest order correction to the current-current correlation
function $\Pi^{12}_{yy}$ in the $|{\bf q}| \rightarrow 0$ limit 
is given by two diagrams in Fig.1. 
The wavy line represents the transverse part of the gauge field
and it can be either ${\bf a}^+$ or ${\bf a}^-$.  The sum of these diagrams
turns out to involve an integral over $(D^+-D^-)^2$.  At small external frequency
$\Omega$ the resulting expression is dominated by the long wavelengths for
which $D^- \gg D^+$.  To get the leading behavior in this limit it is
sufficient to take both wavy lines to involve ${\bf a}^-$ fluctuations and
to evaluate the the resulting expressing by scaling.  The important internal
momenta turn out to be of $q_{\Omega} \sim (\gamma \Omega/\chi^-)^{1/3}$, and
the condition for the validity of the approximation is $q_{\Omega}d \ll 1$;
{\it i.e.} $\Omega \ll \Omega_{cr}$ with
\begin{equation}
\Omega_{\rm cr} = {\chi^- \over \gamma d^3}
\label{omegacr}
\end{equation}
For larger frequencies, the planes become effectively decoupled and the
gauge contribution to the transconductivity drops rapidly.  
We also note that another important energy scale is the composite fermion
energy $\varepsilon_F = k_F^2/(2m^*)$; the theory is only valid for 
$\Omega < \varepsilon_F$.
Putting the definitions and numerical factors into Eq \ref{omegacr} yields
\begin{equation}
{\Omega_{\rm cr} \over \varepsilon_F} \approx 
\left ( {l_B \over d} \right )^3 \left ( {2 \over 3} + 
1.65 {d \over l_B} \right ) \ . 
\end{equation}
Notice that this crossover scale is smaller by a numerical factor of 
order 10 than previous estimate obtained by Bonesteel from a calculation
of composite fermion self-energies\cite{Bonesteel}.

Returning now to the the calculation we find that, in the 
$\Omega < \Omega_{\rm cr}$ limit, the sum of two diagrams 
in the $|{\bf q}| \rightarrow 0$ limit can be written in a 
compact way as
\begin{equation}
\Pi^{12}_{yy} (i\nu) = 
{1 \over 8} \int {d^2 q' \over (2 \pi)^2} {d \nu' \over 2\pi}
D^- ({\bf q'},i\nu') D^- ({\bf q'},i\nu + i\nu')
\left [ \Gamma ({\bf q'};i\nu,i\nu') + 
\Gamma (-{\bf q'};i\nu,-i\nu -i\nu') \right ]^2 \ ,
\end{equation}
where 
\begin{equation}
\Gamma ({\bf q'};i\nu,i\nu') = \int {d^2 k \over (2 \pi)^2}
{d \omega \over 2 \pi} {k_y \over m^*} \left ( {k \over m^*}
\ {\rm sin} \theta_{{\bf k}{\bf q}} \right )^2 
G({\bf k},i\omega) \ G({\bf k},i\omega + i\nu) 
\ G({\bf k} + {\bf q'},i\omega + i\nu + i\nu') 
\end{equation}
and $\theta_{{\bf k}{\bf q}}$ represents the angle
between ${\bf k}$ and ${\bf q}$. $G({\bf k},i\omega) = 
1/(i\omega - \xi_{\bf k})$, where $\xi_{\bf k} = k^2 / 2m^* - \mu$,
is the free fermion Green's function. 

In the limit $q' \ll k_F$,
\begin{eqnarray}
&&\Gamma ({\bf q'};i\nu,i\nu') + \Gamma (-{\bf q'};i\nu,-i\nu -i\nu') \cr
&&= {i k_F \over 4 \pi m^* \nu} {q'_y \over q'} \left [
|\nu'| \sqrt{\left ( {\nu' \over v_F q'} \right )^2 + 1}
- |\nu + \nu'| \sqrt{\left ( {\nu + \nu' \over v_F q'} 
\right )^2 + 1} + {\nu^2 + 2 \nu \nu' \over v_F q'} \right ] \ .
\end{eqnarray}
Using this result, one can calculate $\Pi^{12}_{yy} (i\nu)$.
After the analytic continuation  $i\nu \rightarrow \Omega 
+ i\delta$, $\Pi^{12}_{yy} (\Omega)$ can be obtained as
\begin{equation}
\Pi^{12}_{yy} (\Omega) \approx
{k^2_F \over (m^*)^2}{1 \over \gamma \chi_-} \left [
c_1 {1 \over \gamma \chi_- d} - i c_2 
\left ( {\gamma \Omega \over \chi_-} \right )^{1/3} 
\right ] \ ,
\end{equation} 
where $c_1 = \sqrt{3} / 1024 \pi^4$ and 
$c_2 = [{\rm ln}2 + \beta(3/4)] / 2304 \pi^4$.
Therefore, the real and the imaginary parts of the
transconductivity is given by
\begin{eqnarray}
{\rm Re} \ \sigma^{12} (\Omega) &=& c_2 {k_F^2 / (m^*)^2 \over 
\gamma^{2/3} (\chi^-)^{4/3}} {1 \over \Omega^{2/3}} \ , \cr
{\rm Im} \ \sigma^{12} (\Omega) &=& c_1 {k_F^2 / (m^*)^2 \over 
\chi^- \gamma d} {1 \over \Omega} \ .
\label{transcond}
\end{eqnarray}

Now we are going to show that the above result is consistent with 
a modified Drude formula with a frequency dependent decay rate 
$1 / \tau_D (\Omega)$. The generalized Drude formula can be 
written as
\begin{equation}
\sigma^{12} (\Omega) = 
C{n_e \over m^*}{1 \over -i \Omega +  1 / \tau_D (\Omega)} \ .
\label{drude}
\end{equation}
In the limit $1 / \tau_D (\Omega) \ll \Omega$, Eq \ref{drude} becomes
\begin{eqnarray}
{\rm Re} \ \sigma^{12} (\Omega) &\approx& 
C{n_e \over m^*}{1 / \tau_D (\Omega) \over \Omega^2} \ , \cr
{\rm Im} \ \sigma^{12} (\Omega) &\approx&
C{n_e \over m^*}{1 \over \Omega} \ .
\label{drudeN}
\end{eqnarray}
Comparing Eq.\ref{drudeN} and Eq.\ref{transcond}, 
one can see that 
\begin{equation}
C = {\sqrt{3} \over 256 \pi^3}{1 \over m^* \gamma \chi_- d}
\end{equation}
and 
\begin{equation}
1 / \tau_D (\Omega) = 0.3 \Omega (\Omega / \Omega_{\rm cr})^{1/3} 
\label{dragrate}
\end{equation} 
As noted above, the theory only applies for 
$\Omega < {\rm min} (\Omega_{cr},\varepsilon_F)$; for these frequencies
one sees immediately from Eq \ref{dragrate} that $\Omega \tau_D >1$,
so the result is consistent.

For finite temperatures $T$, one can replace $\Omega$ by $T$ 
in Eq.\ref{dragrate}. 
Now one can infer the DC limit
by assuming that the full $\sigma^{12} (\Omega, T)$ 
is given by Eq.\ref{drude} with $1 / \tau_D (\Omega)$
being replaced by $1 / \tau_D (T)$ and by checking whether it is 
consistent with the result of the optical transconductivity at $T=0$.
The considerations above lead to the conclusion that the finite 
temperature drag rate $1 / \tau_D (T)$ is given by 
\begin{equation}
1 / \tau_D (T) = 0.3 T(T / \Omega_{\rm cr})^{1/3} 
\label{dragrateT}
\end{equation}
We now make numerical estimates for $\Omega_{\rm cr}$ and 
${\rm Re} \ \sigma^{12} (T)$.
For $d/l_B \sim 2$, $\Omega_{\rm cr} \sim 0.5 \varepsilon_F$.
Since $\varepsilon_F \sim 4 meV$, $\Omega_{\rm cr} \sim 2 meV
\sim 23 {\rm K}$. Thus $T^{4/3}$ should be observable for 
$T < 23 {\rm K}$.
${\rm Re} \ \sigma^{12} (T)$ can be written as
\begin{equation}
{\rm Re} \ \sigma^{12} (T) = {e^2 \over \hbar} 
{\sqrt{3} \over 32 \pi^2}{1 \over 4/3 + 3.3 d/l_B}
{l_B \over d} (\varepsilon_F \tau_D) \ . 
\end{equation}
If $d/l_B = 2$, ${\rm Re} \ \sigma^{12} (T) = 1.2 \times 
10^{-3} (\varepsilon_F / T) (\Omega_{\rm cr} / T)^{1/3}$. 
For $T = 1 {\rm K}$, ${\rm Re} \ \sigma^{12} = 0.16 e^2/\hbar$. 

The transresistivity $\rho^{12}$ can be expressed in terms
of the transconductivity $\sigma^{12}$ and the in-plane 
conductivities $\sigma^{11}$ and $\sigma^{22}$ as
$\rho^{12} = \sigma^{12} / (\sigma^{11} \sigma^{22}
- \sigma^{12}\sigma^{21})$\cite{Jari,Oreg}.
Even in the clean limit, $\sigma^{11}$ and $\sigma^{22}$ 
in the drag geometry are finite because the system is
not Galilean-invariant.
Since $\sigma^{11}$ and $\sigma^{22}$ are determined from
the in-plane transport scattering rate which is also 
proportional to $T^{4/3}$\cite{Nagaosa,HLR}, the 
transresistivity is still proportional to $1 / \tau_D$.

In summary, we calculate the drag rate in double-layers
of half-filled Landau levels. The drag is dominated by the 
scattering of composite fermions by the antisymmetric 
combination of the gauge fields or the density fluctuations 
in two layers. It is found that this {\it gauge drag} rate 
is proportional to $T^{4/3}$.

{\it Acknowledgements}: This work was initiated from the
discussion with J. P. Eisenstein and his preliminary
experimental data. We thank him for showing his
unpublished data to us and illuminating discussions. 
We also thank N. E. Bonesteel, A. Furusaki, B. I. Halperin, S. He, 
G. B. Kotliar, P. A. Lee, P. B. Littlewood, A. H. MacDonald,
N. Nagaosa, and X. G. Wen for helpful discussions.

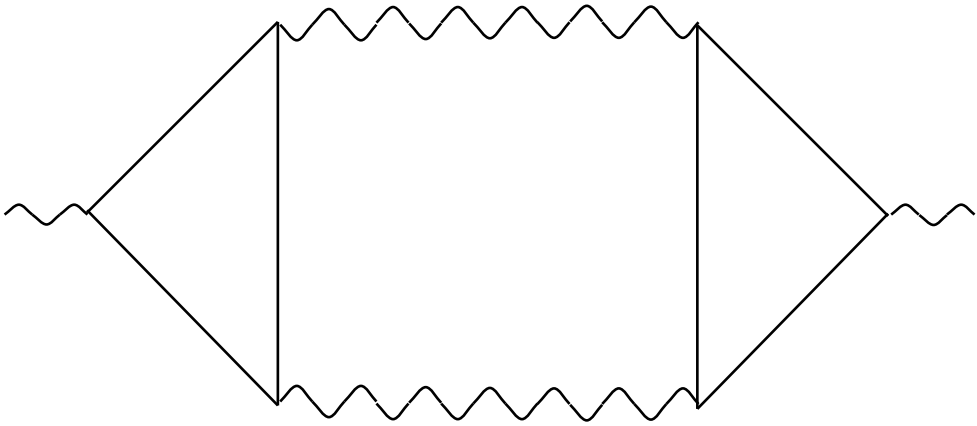
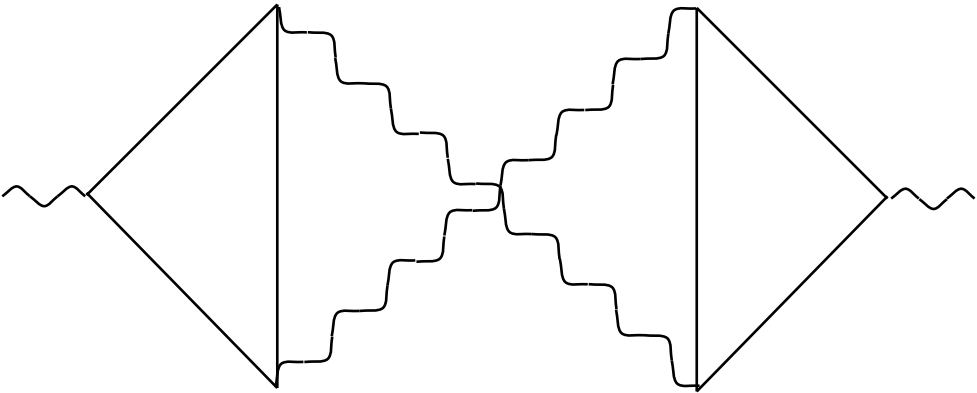
\begin{figure}
\caption{
The leading non-vanishing diagrams which contribute to
the optical transconductivity.
The wavy line corresponds to the gauge field and the
solid line corresponds to the fermions.}
\end{figure}
 
\end{document}